\documentclass[conference]{IEEEtran}

\usepackage{cite}
\usepackage{amsmath,amssymb,amsfonts}
\usepackage{algorithmic}
\usepackage{graphicx}
\usepackage{textcomp}
\usepackage{xcolor}
\def\BibTeX{{\rm B\kern-.05em{\sc i\kern-.025em b}\kern-.08em
    T\kern-.1667em\lower.7ex\hbox{E}\kern-.125emX}}

\DeclareMathOperator{\Tr}{Tr}
\newcommand\numberthis{\addtocounter{equation}{1}\tag{\theequation}} 
\newcommand{\minus}{\scalebox{0.75}[1.0]{$-$}}
\DeclareMathOperator*{\argmax}{arg\,max}

\usepackage{subcaption}
\usepackage{balance}

\begin{document}

\title{Hybrid Fusion for 802.11ax Wi-Fi-based Passive Radars Exploiting Beamforming Feedbacks}
\author{\IEEEauthorblockN{
Martin Willame\IEEEauthorrefmark{1}\IEEEauthorrefmark{2},   
Hasan Can Yildirim\IEEEauthorrefmark{2},   
Laurent Storrer \IEEEauthorrefmark{2},    
François Horlin\IEEEauthorrefmark{2},      
Jérôme Louveaux\IEEEauthorrefmark{1},      
}                                     
\IEEEauthorblockA{\IEEEauthorrefmark{1}
UCLouvain - ICTEAM: Université catholique de Louvain, Louvain-la-Neuve, Belgium 
\IEEEauthorblockA{\IEEEauthorrefmark{2}
ULB - OPERA: Université libre de Bruxelles, Brussels, Belgium}}
\IEEEauthorblockA{\{martin.willame, jerome.louveaux\}@uclouvain.be,\{hasan.can.yildirim, laurent.storrer, francois.horlin\}@ulb.be}}

%

\maketitle

\begin{abstract}
Passive Wi-Fi-based radars (PWRs) are devices that enable the localization of targets using Wi-Fi signals of opportunity transmitted by an access point. Unlike active radars that optimize their transmitted waveform for localization, PWRs align with the 802.11 amendments. Specifically, during the channel sounding session preceding a multi-user multiple-input multiple-output downlink transmission, an access point isotropically transmits a null data packet (NDP) with a known preamble. From these known symbols, client user equipments derive their channel state information and transmit an unencrypted beamforming feedback (BFF) back to the access point. The BFF comprises the right singular matrix of the channel and the corresponding stream gain for each subcarrier, which allows the computation of a beamforming matrix at the access point. In a classical PWR processing, only the preamble symbols from the NDP are exploited during the channel sounding session. In this study, we investigate multiple target localization by a PWR exploiting hybrid information sources. On one hand, the joint angle-of-departure and angle-of-arrival evaluated from the NDP. On another hand, the line-of-sight angle-of-departures inferred from the BFFs. The processing steps at the PWR are defined and an optimal hybrid fusion rule is derived in the maximum likelihood framework. Monte-Carlo simulations assess the enhanced accuracy of the proposed combination method compared to classical PWR processing based solely on the NDP, and compare the localization performance between client and non-client targets.
\end{abstract}

\vskip0.5\baselineskip

\section{Introduction} \label{sec:intro}
Passive Wi-Fi-based Radars (PWRs) are devices employed for target localization utilizing Wi-Fi signals transmitted by an Access Point (AP) \cite{6178049}. The communication signal emitted by the AP to its clients within the environment is reflected off the targets and received by the PWR. The PWR captures and demodulates known preamble orthogonal frequency division multiplexing (OFDM) modulated symbols to estimate the localization parameters of the targets. Unlike active radars which optimize their transmitted waveform for localization, the design of PWRs is intrinsically linked to the evolution of the 802.11 standard \cite{5393298}. Consequently, any wireless communication technology specified within the Wi-Fi standard presents an opportunity to enhance PWR localization capabilities.

In particular, the initiation phase of a multi-user multiple-input multiple-output (MU-MIMO) downlink transmission between an AP and multiple client user equipments (UEs) presents a valuable opportunity. From a communication perspective, an 802.11ax AP initiates a channel sounding session to determine its beamforming coefficients. During this session, null data packets (NDP) containing known OFDM symbols are transmitted to the client UEs. Using the NDP, the client UEs can evaluate their channel state information (CSI) and send unencrypted beamforming feedback (BFF) back to the AP. These MU-MIMO BFFs consist of a subcarrier-averaged stream gain, the right singular matrix of the channel for each subcarrier, and the quantized delta streams, allowing the AP to compute its beamforming matrix \cite{9442429}. From a sensing perspective, while traditional PWR processing during the channel sounding session relies solely on the known preambles provided by the NDP, the BFF can also be leveraged by the PWR to enhance its sensing capabilities.

Previous works in the literature explore the utilization of the BFF for various purposes. For instance, an experimental investigation conducted by \cite{9787542} focuses on BFF-based angle-of-departure (AoD) estimation. The study demonstrates that the estimation error of AoD using BFF is comparable to that achieved through CSI-based estimation methods. In \cite{10107509}, the right singular vectors contained within the BFF are intercepted at a PWR to reconstruct the precoded MIMO preamble signals. This approach enables the localization of targets using the beamformed Wi-Fi signals. However, to the best of the author's knowledge, there has been no investigation into the hybrid combination of the CSI evaluated at a PWR during the channel sounding session with the information provided by the BFF to enhance the localization accuracy of the radar system. Additionally, there is no previous work leveraging the delta signal-to-noise ratio (SNR) transmitted in the MU-MIMO BFF for this purpose.

In this study, our focus lies on the localization of $K$ targets by a PWR exploiting a hybrid source of information: the joint AoDs and angles-of-arrival (AoAs) extracted from the NDP transmitted by an AP, and the line-of-sight (LoS) AoD inferred from the BFF transmitted by the client UEs. Consequently, the PWR must implement a combination rule for the optimal exploitation of both sources of information.

The maximum likelihood (ML) framework enables the derivation of an optimal joint AoD/AoA-based fusion rule. However, in scenarios involving the detection of multiple targets, the complexity of the ML algorithm increases, as its brute force implementation requires a $2K$-dimensional search across the positions of all targets \cite{57542}. In our previous analysis \cite{willame2024EuSIPCO}, we introduced a fusing methodology for a multistatic OFDM radar localization based on known preambles only, effectively decoupling the $2K$-dimensional ML estimator into $K$ per-target two-dimensional searches. The proposed alternating summation method relies on a pre-estimation of the targets parameters acquired via the multiple signal classification (MUSIC) algorithm. This method takes into account the varying ability of the different radar pairs constituting the multistatic configuration to localize different targets.

The aim of this paper is to explore the advantages at a PWR of jointly utilizing channel estimation from the NDP and AoD information derived from the BFFs for localizing multiple targets. Our key contributions are as follows:
\begin{itemize}
\item We outline the processing steps at the PWR for extracting both the CSI from the NDP transmitted by the AP and the BFFs transmitted by the client UEs during the MU-MIMO channel sounding session.
\item We propose the associative alternating summation method for localizing multiple targets through hybrid fusion of the CSI and BFFs. This method extends the alternating summation technique from \cite{willame2024EuSIPCO} by incorporating an association step and calculating an approximate sample covariance matrix of the channel using the right singular vectors and quantized subcarrier stream gain from the BFF.
\item Numerical simulations highlight the benefits of exploiting BFFs over a localization method based solely on CSI. Additionally, we compare localization performance between client and non-client targets.
\end{itemize}
The structure of the paper is organized as follows: Section \ref{sec:CSSS} presents the sensing steps at the PWR during the channel sounding session. Section \ref{sec:system_model} describes the system model. Section \ref{sec:BBF_Radar_Process} details the mathematical formulation of the hybrid radar processing at the PWR. In Section \ref{sec:sim_res}, numerical results are provided to evaluate the benefit of the proposed fusion.

\subsection{Notations}
The vectors and matrices are defined as $\mathbf{a}$ and $\mathbf{A}$, respectively. The trace, the transpose and the Hermitian transpose are denoted by $\Tr\left\{\mathbf{A}\right\} ,\mathbf{A}^{\mathrm T}$ and $\mathbf{A}^{\mathrm H}$, respectively. The Moore-Penrose inverse is defined as $\mathbf{A}^{+} =\left(\mathbf{A}^{\mathrm H} \mathbf{A} \right)^{-1} \mathbf{A}^{\mathrm H}$. The vector and the Frobenius matrix norm are written $\lVert \mathbf{a} \rVert$ and $\lVert \mathbf{A} \rVert_{\mathrm F}$. The identity matrix is denoted by $\mathbf{I}$ and the Kronecker product is denoted by $\otimes$.
\section{Channel Sounding Session Sensing} \label{sec:CSSS}
\begin{figure}
\centering
\includegraphics[width=0.95\columnwidth]{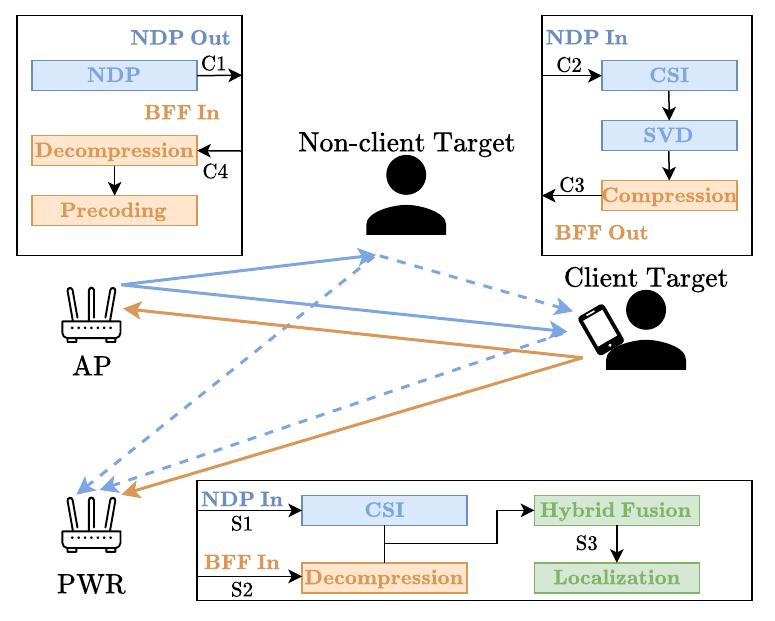}
\caption{Illustration of the scenario. The hybrid configuration comprises an AP, a PWR, multiple client and non-client targets. In the illustration, solid lines represent the incident waveforms, while dashed lines denote the reflected waveforms. Blue lines represent the NDP, and orange lines depict the BFF.}
\label{fig:scenario}
\end{figure}

In this study, we investigate the sensing capability of a PWR during the channel sounding session defined by the 802.11ax amendment. The scenario is illustrated in \figurename \ref{fig:scenario}.

From a communication standpoint, the AP initiates a downlink MU-MIMO transmission with the client UEs. To compute or update its beamforming steering matrix, the AP initiates an explicit channel sounding session with the following steps:
\begin{enumerate}
\item[C1)] The AP isotropically transmits an NDP to the client UEs.
\item[C2)] The client UEs evaluate their CSI for each transmitting-receiving antenna pairing from known high efficiency long training fields (HE-LTF) in the NDP.  
\item[C3)] Each client UE performs a singular value decomposition (SVD) of the obtained channel matrix to extract the right singular matrix of the channel and the quantized delta stream for each subcarrier, along with a subcarrier-averaged stream gain. These parameters are mathematically defined in Section \ref{sec:BBF_Radar_Process}. Subsequently, they are transmitted back to the AP in an unencrypted and compressed BFF.
\item[C4)] The AP decompresses the BFFs and computes its precoding matrix for beamformed data transmission.
\end{enumerate} 

From a sensing perspective, the PWR aims to localize the clients and other potential targets within its coverage area. To achieve this, the PWR operates as follows:
\begin{enumerate}
\item[S1)] It evaluates its CSI from the NDP transmitted isotropically by the AP.
\item[S2)] It intercepts the BFFs transmitted by the client UE and uncompresses their content.
\item[S3)] It combines the estimated CSI with the information contained in the BFFs to enhance the localization of the targets. The procedure for the combination is defined in Section \ref{sec:BBF_Radar_Process}.
\end{enumerate} 
\section{System Model} \label{sec:system_model}
In this section, we provide the system model for the configuration depicted in \figurename~\ref{fig:scenario}. On one hand, the AP initiates a channel sounding session to conduct a downlink MU-MIMO transmission to the client UEs. As described in Section \ref{sec:CSSS}, the AP isotropically transmits an NDP to obtain feedback from the client UEs, from which the beamforming steering matrix can be evaluated. If the AP has $N_A$ antennas, then the NDP comprises $N_A$ known HE-LTF OFDM symbols on $Q$ subcarriers with a subcarrier spacing of $\Delta_f$.

On the other hand, the PWR intends to exploit this NDP to localize $K$ targets within its coverage area, among which $C$ of them are client targets and $K \minus C$ are non-client targets. The positions of these $K$ targets are defined by the vectors $\mathbf{x}=[x_1 \ \dots \ x_K]^{\mathrm T}$ and $\mathbf{y}=[y_1 \ \dots \ y_K]^{\mathrm T}$, where indices $k=1\dots C$ correspond to the positions of the client UEs. The AP, the PWR and the client UEs are all equipped with uniform linear arrays of respectively $N_A$, $N_P$ and $N_u$ antennas, where $u=1\dots C$. For simplicity, we assume that the arrays of the AP and the PWR are oriented towards the coverage area and have half-wavelength spacing. To express the channel model,
we assume that only multipath signals featuring a single reflection on a target significantly impact the observed channel model. In this single-bounce model, signals with multiple reflections are thus omitted \cite{8403681}. This is a classical
assumption for the development of radar processing
algorithms since second order paths would increase the complexity of the channel model.

In the remainder of this section, we describe the radar channel model between the AP and the PWR, as well as the communication channel model between the AP and the client UEs.

\subsection{Passive Radar Channel Model}
To describe the radar channel between the AP and the PWR, we assume that the direct LoS signal, along with clutter contributions, are effectively suppressed from the estimated channel. This can be obtained by exploiting previously transmitted known preamble symbols. The resulting radar channel is thus defined by the $K$ multipath components resulting from reflections on the client and non-client targets. The AoD and the AoA corresponding to the $k\textsuperscript{th}$ target are denoted by $\varphi_k^r$ and $\vartheta_{k}^r$, respectively. All angles are defined between the wavefront and the normal vector of their corresponding antenna array. The corresponding complex steering vectors are thus given by $\mathbf{a}(\varphi_{k}^r)= [1 \ e^{j\pi \sin(\varphi_{k}^r)} \dots \ e^{j\pi (N_A-1) \sin(\varphi_{k}^r)}]^{\mathrm T}, \in \mathbb{C}^{N_A \times 1}$ and $\mathbf{a}(\vartheta_{k}^r)= [1 \ e^{j\pi \sin(\vartheta_{k}^r)} \dots \ e^{j\pi (N_P-1) \sin(\vartheta_{k}^r)}]^{\mathrm T}, \in \mathbb{C}^{N_P \times 1}$. The sets of all $K$ AoDs and AoAs are denoted by $\boldsymbol \Phi^r$ and $\boldsymbol\Theta^r$, respectively. Throughout this paper, for the sake of notation simplicity, we do not explicitly denote the dependence of the angle on $(\mathbf{x},\mathbf{y})$. The baseband equivalent channel matrix for the radar channel is defined for each subcarrier $q$ as follows
\begin{equation}\label{eq:rad_channel}
\mathbf{H}_{q}^{r} = \mathbf{A}(\boldsymbol \Theta^r)\  \mathbf{B}_{q}^{r} \ \mathbf{A}^{\mathrm H}(\boldsymbol \Phi^r), \ \in \mathbb{C}^{N_P \times N_A},
\end{equation}
where $\mathbf{A}(\boldsymbol\Phi^r) = [\mathbf{a}(\varphi_1^r) \ \dots \ \mathbf{a}(\varphi_K^r)] \in \mathbb{C}^{N_A \times K}$ is the AoD steering matrix, $\mathbf{A}(\boldsymbol\Theta^r) = [\ \mathbf{a}(\vartheta_1^r) \ \dots \ \mathbf{a}(\vartheta_K^r)] \in \mathbb{C}^{N_P \times K}$ is the AoA steering matrix, and $\mathbf{B}_{q}^{r} \in \mathbb{C}^{K \times K}$ is the radar channel coefficient diagonal matrix. The diagonal entries of the matrix are defined by the channel vector $\boldsymbol \beta_{q}^{r} = [\beta_{q,1}^{r} \ \dots \ \beta_{q,K}^{r}]^{\mathrm T}, \in \mathbb{C}^{K \times 1}$. The received signal model at the PWR is thus given by
\begin{equation}\label{eq:rad_received}
\mathbf{Z}_{q}^{r} = \mathbf{H}_{q}^{r} \mathbf{S}_q + \mathbf{N}_{q}^{r},
\end{equation}
where $\mathbf{N}_{q}^{r} \in \mathbb{C}^{N_P \times N_A}$ denotes the radar Additive White Gaussian Noise (AWGN) matrix, and $\mathbf{S}_q \in \mathbb{C}^{N_A \times N_A}$ is the transmitted signal per subcarrier. The columns of $\mathbf{S}_q$ represent the different HE-LTF OFDM symbols, and the rows represent the different AP antennas. Both the client UE and the PWR know the structure of the transmitted $\mathbf{S}_q$ matrix from the information contained in the preamble of the NDP, as it is well defined in the 802.11ax amendment \cite{9442429}.

\subsection{Communication Channel Model} \label{sec:comm_model}
The communication channel between the AP and the $u\textsuperscript{th}$ client UE consists of the LoS link and $K_u \minus 1$ multipath components resulting from reflections on other targets and the clutter. In this scenario, the clutter contributions cannot be effectively removed from the communication channel, as the client UE does not perform any suppression before computing its BFF. However, it is assumed that the LoS link is significantly stronger, corresponding to a high Ricean K-factor scenario \cite{1210745}. The LoS AoD is denoted by $\varphi_u^c$ and the set $\boldsymbol \Phi_u^c$ comprises all multipath AoDs including the reflection on both the other $K-1$ targets and the clutter. The baseband equivalent channel matrix for the communication channel of the $u\textsuperscript{th}$ client UE is defined for each subcarrier $q$ as follows
\begin{equation} \label{eq:com_channel}
\mathbf{H}_{u,q}^{c} = \boldsymbol \beta_{u,q}^{c} \mathbf{a}^{\mathrm H}(\varphi_u^c) + \mathbf{B}_{u,q}^{c} \ \mathbf{A}^{\mathrm H}(\boldsymbol \Phi_u^c), \in \mathbb{C}^{N_u \times N_A},   
\end{equation}
where $\boldsymbol \beta_{u,q}^{c} \in \mathbb{C}^{N_u \times 1}$ and $\mathbf{B}_{u,q}^{c} \in \mathbb{C}^{N_u \times (K_u-1)}$ represent the LoS and multipath communication channel coefficients, respectively. Each column represents a different path, while the rows represent the different antennas of the UE. These coefficients encompass the linearly increasing phases across subcarriers and across receiving antennas, attributed to the range and AoA of the paths, respectively, as well as the attenuation defined by the radar range equation. The received signal model at the client UE is given by
\begin{equation}\label{eq:com_received}
\mathbf{Z}_{u,q}^{c} = \mathbf{H}_{u,q}^{c} \mathbf{S}_q + \mathbf{N}_{u,q}^{c},
\end{equation}
where $\mathbf{N}_{u,q}^{c} \in \mathbb{C}^{N_u \times N_A}$ denotes the AWGN matrix.

\section{Hybrid Radar Processing}\label{sec:BBF_Radar_Process}
In this section, we present the hybrid radar processing at the PWR used to determine the $(x,y)$ positions of the $K$ targets based on the intercepted BFFs and the NDP processing. First, we derive the ML combination rule using the received signals from the radar channel in \eqref{eq:rad_received} and the communication channels in \eqref{eq:com_received}. It is demonstrated that target localization depends solely on the covariance matrices of the radar and communication channels. Next, we detail the estimation process for the covariance matrix of the communication channel using the BFFs transmitted by each client. Finally, we introduce the associative alternating summation method to reduce the computational complexity of the maximum likelihood estimator. This novel approach extends the alternating summation method from \cite{willame2024EuSIPCO} by incorporating an association step.

\subsection{Maximum Likelihood Fusion} \label{sec:MLF}
As described in Section \ref{sec:system_model}, the PWR and client UEs utilize the HE-LTF symbols transmitted isotropically by the AP to estimate the radar and communication channel matrices, respectively. The CSI matrices at the PWR and client UEs are obtained by multiplying the received signals in \eqref{eq:rad_received} and \eqref{eq:com_received} by $\mathbf{S}_q^{\mathrm H}$. Given that $\mathbf{S}_q \mathbf{S}_q^{\mathrm H} = \mathbf{I}$ and defining $\mathbf{N}_{q}^{r\prime}$ and $\mathbf{N}_{u,q}^{c \prime}$ as the noise matrices with variances $\sigma_r^2$ and $\sigma_u^2$ after symbol equalization, the estimated CSI matrices can be expressed as
\begin{align*}
\widetilde{\mathbf{H}}_{q}^{r} & = \mathbf{Z}_{q}^{r} \mathbf{S}_q^{\mathrm H}  = \mathbf{H}_{q}^{r} + \mathbf{N}_{q}^{r\prime}, \numberthis \label{eq:obs_r} \\
\widetilde{\mathbf{H}}_{u,q}^{c} & = \mathbf{Z}_{u,q}^{c} \mathbf{S}_q^{\mathrm H}  = \mathbf{H}_{u,q}^{c} + \mathbf{N}_{u,q}^{c\prime}. \numberthis\label{eq:obs_c}
\end{align*}
To localize the $K$ targets in the scene, the PWR exploits a hybrid source of information: the joint AoD/AoA extracted from the NDP and the LoS AoD from the BFFs. This is highlighted in the ML formulation by rewriting the estimated CSI matrices \eqref{eq:obs_r} and \eqref{eq:obs_c} as
\begin{align*}
\widetilde{\mathbf{h}}_{q}^{r} & = \mathbf{A}^\prime(\boldsymbol \Phi^r,\boldsymbol \Theta^r) \ \boldsymbol \beta_{q}^{r} + \mathbf{n}_{q}^{r\prime}, \numberthis \label{eq:CSI_r} \\
\widetilde{\mathbf{H}}_{u,q}^c & = \boldsymbol \beta_{u,q}^{c} \mathbf{a}^{\mathrm H}(\varphi_u^c) + \mathbf{N}_{u,q}^{c \prime}, \numberthis \label{eq:CSI_c}
\end{align*}
where $\widetilde{\mathbf{h}}_{q}^{r}, \mathbf{n}_{q}^{r\prime} \in \mathbb{C}^{N_A N_P \times 1}$ are the vector obtained by vectorizing the radar CSI matrix $\widetilde{\mathbf{H}}_{q}^{r}$ and the radar noise matrix $\mathbf{N}_{q}^{r\prime}$, respectively, and where $\mathbf{A}^\prime(\boldsymbol \Phi^r, \boldsymbol \Theta^r)=[\mathbf{a}^\prime(\varphi_1^r,\vartheta_1^r) \ \dots \ \mathbf{a}^\prime(\varphi_K^r,\vartheta_K^r)] \in \mathbb{C}^{N_A N_P \times K}$ represents the joint AoD/AoA matrix in which $\mathbf{a}^\prime(\varphi_k^r,\vartheta_k^r) = \mathbf{a}(\varphi_k^r) \otimes \mathbf{a}(\vartheta_k^r), \mathbb{C}^{N_A N_P \times 1}$.

Observe that in \eqref{eq:CSI_c}, the multipath term $\mathbf{B}_{u,q}^{c} \ \mathbf{A}^{\mathrm H}(\boldsymbol \Phi_u^c)$ is omitted from the model of the observations as only the LoS AoD to the client UE will be retrieved from the BFFs. The reasoning for this model mismatch is further discussed in Section \ref{sec:bff}.

The parameters to be estimated are defined by the vector $\boldsymbol \gamma = [\mathbf{x}^{\mathrm{T}} \mathbf{y}^{\mathrm{T}} \{\boldsymbol\beta_{q}^{r}\}_{q=1\dots Q} \{\boldsymbol\beta_{u,q}^{c}\}_{q=1\dots Q}^{u=1\dots C}]^{\mathrm{T}}$. It is assumed that the number of targets observed by the PWR is known, as methods for estimating $K$ are available in the literature \cite{1164557}. Notice that our study solely focuses on angle-based ML localization and does not utilize range information. Therefore, each channel coefficient vector $\boldsymbol\beta_{q}^{r}, \boldsymbol\beta_{u,q}^{c}$ is independently estimated, as the linear phase increase across subcarriers, defined by the range of each target, is not exploited. Considering independent noise contributions for the estimated CSIs, the combined likelihood function is derived as the product of individual Gaussian density functions. After taking the natural logarithm of the combined likelihood function, the maximization problem can be written as
\begin{equation} \label{eq:global_max}
\widehat{\boldsymbol \gamma} = \argmax_{\boldsymbol \gamma} \mathcal{L}(\boldsymbol \gamma) = \argmax_{\boldsymbol \gamma} \mathcal{L}^r(\boldsymbol \gamma) + \sum_{u=1}^{C} \mathcal{L}_u^c(\boldsymbol \gamma),
\end{equation}
where the log-likelihood from the PWR and the clients CSIs are respectively given by
\begin{align*}
\mathcal{L}^r(\boldsymbol \gamma) & = \frac{-1}{2\sigma_r^2}\sum_{q=1}^{Q} \lVert \widetilde{\mathbf{h}}_{q}^{r} - \mathbf{A}^{\prime}(\boldsymbol \Phi^r,\boldsymbol \Theta^r) \ \boldsymbol \beta_{q}^{r} \rVert^2, \numberthis \label{eq:log_like_r}\\
\mathcal{L}_u^c(\boldsymbol \gamma) & = \frac{-1}{2\sigma_u^2} \sum_{q=1}^{Q} \lVert \widetilde{\mathbf{H}}_{u,q}^{c}- \boldsymbol \beta_{u,q}^{c} \mathbf{a}^{\mathrm H}(\varphi_u^c)\rVert_{\mathrm F}^{2}, \numberthis \label{eq:log_like_c}
\end{align*}

First, we maximize with respect to the channel coefficients $\boldsymbol\beta_{q}^r$ and $\boldsymbol\beta_{u,q}^{c}$ to obtain closed-form expressions as function of the angles. Following the solution of the resulting linear least squares problem, the ML estimate of the channel coefficients for every subcarrier $q$ is expressed as $\widehat{\boldsymbol\beta}_{q}^r(\boldsymbol \Phi^r,\boldsymbol \Theta^r) = \left(\mathbf{A}'(\boldsymbol \Phi^r,\boldsymbol \Theta^r)\right)^{+}\ \widetilde{\mathbf{h}}_{q}^r$ and $\widehat{\boldsymbol\beta}_{u,q}^c(\varphi_u^c) = \widetilde{\mathbf{H}}_{u,q}^c \left(\mathbf{a}^{+}(\varphi_u^c)\right)^{\mathrm H}$.

Substituting these estimates  back into \eqref{eq:log_like_r} and \eqref{eq:log_like_c}, and after performing some mathematical manipulations, the two log-likelihood functions can be reformulated as 
\begin{align*}
\mathcal{L}^r(\boldsymbol \gamma) & = \frac{Q}{2\sigma_r^2} \Tr\left\{ \mathbf{A}'(\boldsymbol \Phi^r,\boldsymbol \Theta^r)\left(\mathbf{A}'(\boldsymbol \Phi^r,\boldsymbol \Theta^r)\right)^{+} \ \widetilde{\mathbf{R}}^r\right\}, \numberthis \label{eq:trace_log_likelihood_r} \\
\mathcal{L}_u^c(\boldsymbol \gamma) & = \frac{Q N_u}{2\sigma_u^2} \Tr\left\{\mathbf{a}(\varphi_u^c)\mathbf{a}^{+}(\varphi_u^c)  \ \widetilde{\mathbf{R}}_u^c\right\}, \numberthis \label{eq:trace_log_likelihood_c}
\end{align*}
in which the sample covariance matrices of the channel averaged over the subcarriers are given by
\begin{align*}
\widetilde{\mathbf{R}}^r & = \frac{1}{Q} \sum_{q=1}^{Q} \widetilde{\mathbf{h}}_{q}^r \ (\widetilde{\mathbf{h}}_q^{r})^{\mathrm H}. \numberthis \label{eq:sample_cov_r} \\
\widetilde{\mathbf{R}}_u^c & = \frac{1}{Q N_u} \sum_{q=1}^{Q} ( \widetilde{\mathbf{H}}_{u,q}^{c})^{ \mathrm H} \ \widetilde{\mathbf{H}}_{u,q}^c, \numberthis \label{eq:sample_cov_c} 
\end{align*}
We have demonstrated that the maximum likelihood estimator depends solely on the knowledge at the PWR of the covariance matrices for both the radar channel and each client's communication channels.

\subsection{Estimation of the covariance matrices} \label{sec:bff}
As outlined in Section \ref{sec:CSSS}, the PWR does not have access to the full CSI, $\widetilde{\mathbf{H}}_{u,q}^c$, estimated by the client UEs. Consequently, while the PWR can compute $\widetilde{\mathbf{R}}^r$ from \eqref{eq:sample_cov_r}, it is unable to directly estimate $\widetilde{\mathbf{R}}_u^c$ from \eqref{eq:sample_cov_c}. However, we demonstrate that the PWR can still approximate the clients' covariance matrices using the intercepted BFFs.

Each client UE generates this feedback by performing an SVD on the estimated CSI matrix, resulting in:
\begin{equation} \label{eq:SVD}
\widetilde{\mathbf{H}}_{u,q}^c = \widetilde{\mathbf{U}}_{u,q}^c \widetilde{\boldsymbol \Sigma}_{u,q}^c (\widetilde{\mathbf{V}}_{u,q}^{c})^{\mathrm H},
\end{equation}
where $\widetilde{\mathbf{U}}_{u,q}^c$ and $\widetilde{\mathbf{V}}_{u,q}^c$ represent the left and right unitary singular matrices, respectively, and $\widetilde{\boldsymbol \Sigma}_{u,q}^c$ is the singular values diagonal matrix. 

As defined in the 802.11ax standard \cite{9442429}, the MU-MIMO BFF contains for each subcarrier:
\begin{itemize}
\item the compressed right singular vector of $\widetilde{\mathbf{V}}_{u,q}^c$ corresponding to the strongest singular value. This vector is denoted $\widetilde{\mathbf{v}}_{u,q}^{c,1}$. 
\item the quantized strongest singular value, denoted $\widetilde{ \Sigma}_{u,q}^{c,1}$, representing the channel gain for the strongest path. In practice, this value is deducted from the subcarrier-averaged stream gain and the delta SNR per subcarrier contained in the BFF. 
\end{itemize}
Details on the compression can be found in \cite{9442429}. As defined in Section \ref{sec:comm_model}, the LoS link between the AP and the client UE is assume to be significantly stronger than the multipath components. In this high Ricean K-factor scenario, the communication covariance matrix $\widetilde{\mathbf{H}}_{u,q}^c$ is defined by the LoS path. Therefore, the sample covariance matrix in \eqref{eq:sample_cov_c} can be approximated from the BFF as follows:
\begin{equation} \label{eq:sample_cov_c_approx}
\widetilde{\mathbf{R}}_u^c \approx \frac{1}{Q N_u} \sum_{q=1}^{Q} \widetilde{\mathbf{v}}_{u,q}^{c,1} \widetilde{\Sigma}_{u,q}^{c,1} (\widetilde{\mathbf{v}}_{u,q}^{c,1})^{\mathrm H}.
\end{equation}
The PWR has thus access to a quantized version of the covariance matrices for the LoS link between the AP and each of the clients.

\subsection{Alternating summation with association step} \label{sec:ASAS}
In this section, we describe the successive steps of the associative alternating summation method, which adapts the approach presented in \cite{willame2024EuSIPCO} to the hybrid fusion scenario under consideration. As discussed in Section \ref{sec:MLF}, our objective is to estimate the $(x, y)$ positions of the $K$ targets by maximizing the sum of the log-likelihood functions for both the radar channel in \eqref{eq:log_like_r} and the client channels in \eqref{eq:log_like_c}. Consequently, the brute-force maximization of the sum in \eqref{eq:global_max} results in a $2K$-dimensional problem due to the presence of multiple targets within the coverage area. In our previous work \cite{willame2024EuSIPCO}, we demonstrated that when an initial estimation of the target parameters is available from each information source, the $2K$-dimensional log-likelihood function can be approximately decoupled into $K$ two-dimensional functions, one per target. 

This pre-estimation can be derived from the radar sample covariance matrix by computing the two-dimensional MUSIC spectrum, expressed as
\begin{equation} \label{eq:MUSIC spectrum}
\mathcal{J}^r(\varphi^r,\vartheta^r) = \frac{1}{(\mathbf{a}^\prime(\varphi^r,\theta^r))^\textup{H}~ \widetilde{\mathbf{G}}^r (\widetilde{\mathbf{G}}^r)^\textup{H} ~\mathbf{a}^r(\varphi^r,\theta^r)},
\end{equation}
where $\widetilde{\mathbf{G}}^r$  represents the estimated noise subspace matrix obtained from the singular value decomposition of $\widetilde{\mathbf{R}}^r$. The $K$ eigenvectors corresponding to the strongest eigenvalues form the signal subspace, while the remaining vectors form the noise subspace \cite{1143830}. The $K$ highest peaks in $\mathcal{J}^r(\varphi^r,\vartheta^r)$, denoted as ($\Hat{\varphi}_{k}^r,\Hat{\theta}_{k}^r$) for each target $k$, constitute the pre-estimated set of joint AoD/AoA values.

Similarly, the LoS AoD for each client can be pre-estimated from their respective approximated sample covariance matrix $\widetilde{\mathbf{R}}_u^c$, as defined in \eqref{eq:sample_cov_c_approx}. For each client $u = 1, \dots, C$, this pre-estimation, denoted as $\widehat{\varphi}_u^c$, corresponds to the highest peak in
\begin{equation}
\mathcal{J}_u^c(\varphi^c) = \mathbf{a}^\textup{H}(\varphi^c)~ \widetilde{\mathbf{R}}_u^c  ~\mathbf{a}(\varphi^c).
\end{equation}
Next, the AoDs $\widehat{\varphi}_u^c$ extracted for each client are associated with the AoDs $\widehat{\varphi}_{k}^r$ estimated from the radar channel. This type of assignment problem is well-known in the literature \cite{9781611972238}. In this work, the association is performed using the Hungarian algorithm \cite{Kuhn2010}. After this step, the information available at the PWR for each target $k$ is represented by the vector $\widehat{\boldsymbol\psi}_k$, as defined in
\begin{equation}
    \Hat{\boldsymbol\psi}_k=\begin{cases}
  \multicolumn{1}{@{}c@{\quad}}{[\widehat{\varphi}_{k}^r,\widehat{\vartheta}_{k}^r,\widehat{\varphi}_u^c],} & \text{ if }\widehat{\varphi}_{u}^c\leftarrow \widehat{\varphi}_{k}^r,\\ 
  \multicolumn{1}{@{}c@{\quad}}{[\widehat{\varphi}_{k}^r,\widehat{\vartheta}_{k}^r],} & \text{ if }\widehat{\varphi}_{u}^c\nleftarrow \widehat{\varphi}_{k}^r ~\forall u, \\
\end{cases}
\end{equation}
where $\leftarrow$ and $\nleftarrow$ indicate an association and  no association, respectively. The complete set of pre-estimated parameters is defined as $\widehat{\boldsymbol\Psi}=\{(\widehat{\boldsymbol\psi}_k)_{k=1\dots K}\}$. As describe in \cite{willame2024EuSIPCO} and \cite{7543}, these pre-estimations allow the decoupling of the $2K$-dimensional problem in \eqref{eq:global_max} into $K$ two-dimensional problems.  Specifically, to localize the $k\textsuperscript{th}$ target, the parameters of the remaining $K-1$ targets are fixed to their values in $\widehat{\boldsymbol\Psi}$, and the corresponding log-likelihood is maximized with respect to $\mathbf{p}_k = (x_k, y_k)$ only. Mathematically, the position $\mathbf{p}_k$ of the $k\textsuperscript{th}$ target is obtained by solving the following maximization problem:
\begin{equation} \label{eq:pertarget}
    \mathbf{p}_k=\argmax_{\mathbf{p}_k} \begin{cases}
  \multicolumn{1}{@{}c@{\quad}}{\mathcal{L}_k^r(\mathbf{p}_k) + \mathcal{L}_k^c(\mathbf{p}_k)} & \text{ if } \widehat{\varphi}_{u}\leftarrow \widehat{\varphi}_{k},\\ 
  \multicolumn{1}{@{}c@{\quad}}{\mathcal{L}_k^r(\mathbf{p}_k)} & \text{ if }\widehat{\varphi}_{u}^c\nleftarrow \widehat{\varphi}_{k}^r ~\forall u. \\
\end{cases}
\end{equation}
The pre-estimation provided by the MUSIC algorithm, which enables the evaluation of \eqref{eq:pertarget}, effectively replaces the $2K$-dimensional ML parameter estimation.
\section{Simulation Results} \label{sec:sim_res}

\begin{figure*}[!t]
\centering
\subfloat[RMSE]{\includegraphics[trim=3.4cm 9.5cm 4.3cm 9.91cm,width = 0.42\textwidth]{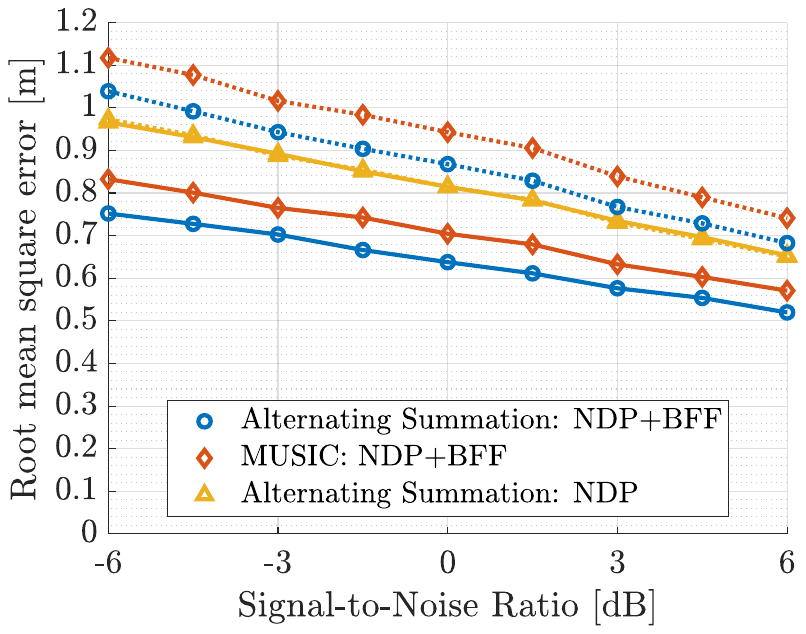}}
\label{fig:graphe1a}
\hfill
\subfloat[Hit Rate]{\includegraphics[trim=3.4cm 9.5cm 4.3cm 9.91cm,width = 0.42\textwidth]{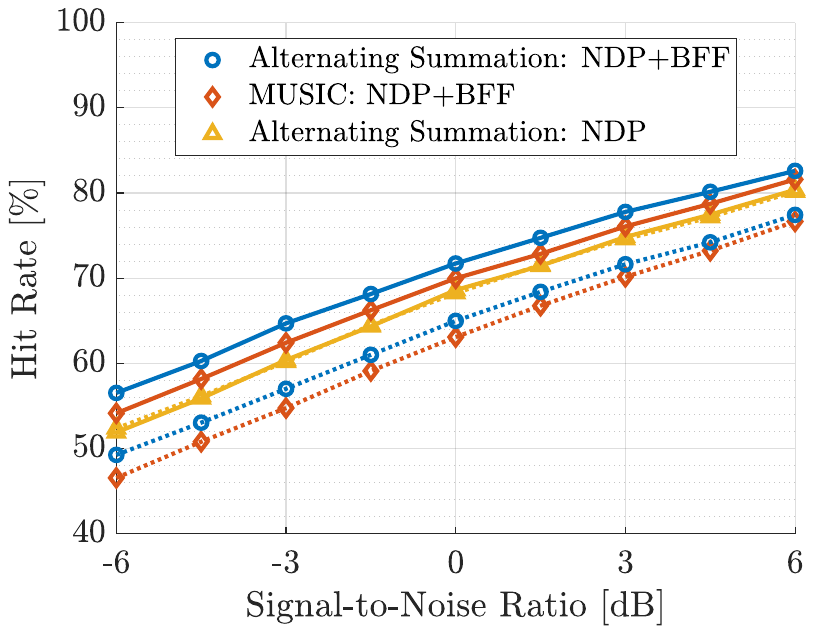}}
\label{fig:graphe1b}
\caption{The impact of the signal-to-noise ratio on the performance metrics of the various methods is depicted. The solid and dashed lines correspond to the localization performance for the client and non-client targets, respectively.}
\label{fig:graphe1}
\end{figure*}
This section evaluates the enhancement brought by the exploitation of the BFFs at the PWR on the accuracy of the radar in localizing targets within its coverage area. The joint NDP and BFF processing outlined in Section \ref{sec:BBF_Radar_Process} is compared through $30,000$ simulations with two different methods:
\begin{enumerate}
\item \textbf{MUSIC combination of NDP and BFF:} In this method, the pre-estimated target positions are refined by replacing the AoD detected by the NDP with that detected by the associated BFF.
\item \textbf{Alternating summation based on NDP only:} This method uses only the information from the NDP, without exploiting the additional BFF data. Target positions are determined using the second line of \eqref{eq:pertarget}.
\end{enumerate}
To compare these methods, the following performance metrics
are studied:
\begin{enumerate}
\item \textbf{Hit Rate:} A hit is acknowledged when a peak is detected within a vicinity of 2 meters from the true target position; otherwise, it is considered a miss.
\item \textbf{RMSE:} The root mean square error (RMSE) is computed between the true target position and the corresponding detected position. To ensure a fair comparison between the methods, only the targets resulting in a hit by all methods are considered in the evaluation of the RMSE.
\end{enumerate}

\balance For each simulation, we analyze a fixed AP and PWR setup designated to locate two client targets and a non-client target randomly positioned. The number of antennas are $N_A=N_u=N_P=4$ and the number of subcarriers is $Q=512$.

\figurename~\ref{fig:graphe1} depicts the results obtained for the hit rate and the RMSE for the client and non client targets as functions of the SNR. The SNR is defined here as the mean of the quotient of the radar channel coefficients of each path by the noise variance. Note that the same noise variance is considered at the clients and the PWR. 

For client targets, the hybrid fusion of NDP and BFF significantly improves RMSE compared to using the NDP alone, with even greater enhancement observed when employing the associative alternating summation method. Regarding the hit rate for client targets, the associative alternating summation method also shows substantial improvement.

However, for non-client targets, both the hit rate and RMSE are slightly negatively impacted by the hybrid combination, as the association step is not always performed accurately. Improving this step could potentially allow the method's performance to match that of the NDP-only algorithm. Overall, these results demonstrate the benefit of hybrid fusion, provided that the fusion rule is appropriately selected.

\section{Conclusion} \label{sec:conclusion}
In this paper, we investigate the multitarget localization capabilities of a PWR during the MU-MIMO channel sounding session initiated by an AP. Both the joint AoD/AoA extracted from the NDP transmitted by the AP and the LoS AoD extracted from the BFFs transmitted by the client UEs are explored for their potential utility by the PWR. The proposed hybrid radar fusion method is derived from the maximum likelihood framework. It is shown to solely rely on the computation, for all source of information, of the approximated sample covariance matrices of the channel state information. Numerical simulations presented in this study validate the effectiveness of the proposed approach. The results highlight the benefits of exploiting BFF in a hybrid fusion for target localization compared to classical PWR processing based solely on NDPs. In future works, the exploitation of range information from the NDP would further enhance localization accuracy.

\bibliographystyle{IEEEtran}
\bibliography{IEEEabrv,References}
\end{document}